\documentclass[epjCONF,columns]{svjour} 
\usepackage{graphicx}
\usepackage[varg]{txfonts} 
\usepackage[latin1]{inputenc}
\session-title{Hadron Collider Physics symposium (HCP-2011)}
\begin{document}
\title{High-p$_t$ in heavy ion collisions: an abridged theoretical overview}
\author{Jos\'e Guilherme Milhano\inst{1}\fnmsep\inst{2}\fnmsep\thanks{\email{guilherme.milhano@ist.utl.pt}}}
\institute{CENTRA, Instituto Superior T\'ecnico, Universidade T\'ecnica de Lisboa, Av. Rovisco Pais, P-1049-001 Lisboa, Portugal \and Physics Department, Theory Unit, CERN, CH-1211 Gen\`eve 23, Switzerland}
\abstract{This overview focusses on recent developments, in the most part triggered by LHC data, aimed at the development of a reliable and complete theoretical description of high-p$_t$ physics in heavy ion collisions. Particular emphasis is placed on the understanding of the underlying in-medium dynamics as a prior to the use of high-p$_t$  observables as detailed probes of the QCD matter created in the collisions.
} 
\maketitle
\section{Introduction}
\label{sec:intro}

Ultra relativistic collisions of heavy ions provide an unique setting where novel properties of QCD can be unveiled and, arguably, understood from first principles. 
The hot and dense QCD matter (a quark-gluon plasma?) created in the collisions manifests itself both in observable collective behaviour (see \cite{jy} for an overview) and in the modification of the QCD dynamics of high-$p_t$  partons that traverse, and thus interact with, the surrounding QCD medium.

The observable consequences of these latter effects, generally referred to as \textit{jet quenching}, have the potential to provide detailed information on the medium properties. 
The first observations of jet quenching at RHIC --- a strong suppression, relatively to what would be expected in the absence of a medium, of the yield of leading high-$p_t$ hadrons; the decrease of the suppression for increasingly peripheral collisions; and the suppression of back-to-back di-hadrons in nucleus-nucleus collisions but not in the deuteron-nucleus control case --- lend support to a picture in which the interaction of the propagating parton with the medium results in significant loss of parton energy. 
While the importance of these measurements cannot be overstated, they provide insufficient information to fully constrain the underlying dynamics. As such, diverse models capable of accounting for RHIC data resulted in large uncertainties for the extracted medium properties \cite{Armesto:2011ht}.

The large jump in centre-of-mass energy afforded by the LHC (2.76 TeV per nucleon pair as compared to 200 GeV at RHIC) and much improved detection capabilities (from the ALICE, ATLAS and CMS detectors) grant access to measurements in an extended kinematic range, importantly well into a domain where the use of perturbative methods can be solidly argued, and also to a whole new class of observables, namely those involving fully reconstructed jets.

This new experimental reality poses a series of challenges to the pre-LHC theory of jet quenching and concurrently offers the necessary support for its further development. While pre-LHC efforts were, for the most part, concerned with the description of energy loss of leading partons and its consequences for observables involving the related leading hadrons, the new theory must take the form of a complete description of the \textit{in-medium} parton shower: jet calculus rules accounting for medium induced modifications of the interference pattern between successive branchings and including parton mass effects; interplay between medium induced radiation and elastic energy loss and related dynamics of radiated soft quanta; a description where a given branching can be ascertained to occur in or outside the medium; understanding of medium recoil;  and the role of colour exchanges between propagating partons and the medium on the subsequent hadronization dynamics.
Once such a detailed understanding of the dynamics of the interactions between parton and medium has been reached, the role of high-$p_t$ observables as medium probes can finally be realized.

 This overview, non-comprehensive and personally biased, focus on recent theoretical efforts that attempt to advance the formulation of a new theory of jet quenching. 

\section{Parton energy loss}
\label{sec:jetcalc}

The calculation of medium-induced parton energy loss, the cornerstone of the theory of jet quenching, has been carried out in a variety of approaches. 

At present, none of these approaches treats parton-medium interactions in rigorous field theoretical terms, with different perturbative QCD based calculations differing on essential points: approximations in the treatment of medium induced parton branching, role of elastic (collisional) energy loss, assumptions regarding the modeling of the medium, and the treatment of multiple gluon emission. These efforts are well documented (see reviews\cite{Wiedemann:2009sh,d'Enterria:2009am,CasalderreySolana:2007zz,Majumder:2010qh,Jacobs:2004qv,Gyulassy:2003mc}), and a systematic comparison has been also carried out \cite{Armesto:2011ht}.  A new approach, based on soft collinear effective field theory, in which many of the kinematical assumptions can be relaxed, has been pursued in \cite{Idilbi:2008vm,D'Eramo:2010ak,D'Eramo:2011zz}.
Further, the effect of strongly coupled media has been elucidated within the context of the AdS/CFT correspondence (for a review see \cite{CasalderreySolana:2011us}).

Some generic features of parton energy loss calculations can be illustrated in rather simple terms. The medium, of extent $L$, is modeled as a discrete set of scatterers and described by a single parameter $\hat{q} = \mu^2/\lambda$ where $\mu^2$ is the squared average momentum transfer from medium to parton on a single scattering, and $\lambda$ the parton mean free path in the medium. The propagating parton is assumed to be sufficiently energetic for its trajectory not to be disturbed by interaction with the medium (eikonal approximation). 
Medium induced radiation occurs when a gluon in the wave function of the propagating parton accumulates sufficient transverse momentum (squared) $\langle k_\perp^2  \rangle \sim \hat{q} L$ to decohere from the parent parton. Thus, the average phase accumulated by the gluon
\begin{equation}
\label{phase}
	\Big\langle \frac{k_\perp^2 L}{\omega}\Big\rangle \sim \frac{\hat{q} L^2}{\omega} \sim \frac{\omega_c}{\omega} \, ,
\end{equation}
should be of order one for emission to take place.
Here, $\omega_c$ is the characteristic gluon energy that sets the scale for the radiated energy distribution.

The shape of the radiated energy distribution can be estimated by considering the number of coherent scatterings that contribute to (\ref{phase}):
\begin{equation}
	N_{coh}\sim \frac{t_{coh}}{\lambda}
\end{equation}
with 
\begin{equation}
	t_{coh} \sim \frac{\omega}{k_\perp^2} \sim \sqrt{ \frac{\omega}{\hat{q}}}\, , \qquad (k_\perp^2\sim \hat{q} t_{coh})\, .
\end{equation}
Then, the radiated energy distribution, per unit length, can be written
\begin{equation}
	\omega\frac{dI_{med}}{d\omega dz} \sim \frac{1}{N_{coh}} \omega \frac{dI_{1 scat}}{d\omega dz} \sim \frac{\alpha_s}{t_{coh}}\sim \alpha_s   \sqrt{ \frac{\hat{q}}{\omega}}\, ,
\end{equation}
where the dependence on $\omega$ is characteristic of the non-abelian Landau-Pomeranchuk-Migdal effect. The average energy loss, with a non-trivial characteristic dependence on $L^2$ follows
\begin{equation}
	\Delta E = \int_0^L dz \int_0^{\omega_c} \omega \frac{dI_{med}}{d\omega dz} \sim \alpha_s \omega_c \sim  \alpha_s \hat{q} L^2 \, .
\end{equation}

\section{Multiple gluon emission}
\label{sec:mult}

Although multiple gluon emission has been considered within several analytical approaches and inevitably in Monte Carlo implementations (an overview can be found in \cite{Zapp:2011kj}), its bona fide implementation requires the understanding of the medium effect on the interference pattern between successive emissions. 

\begin{figure}
\includegraphics[width=0.48\textwidth]{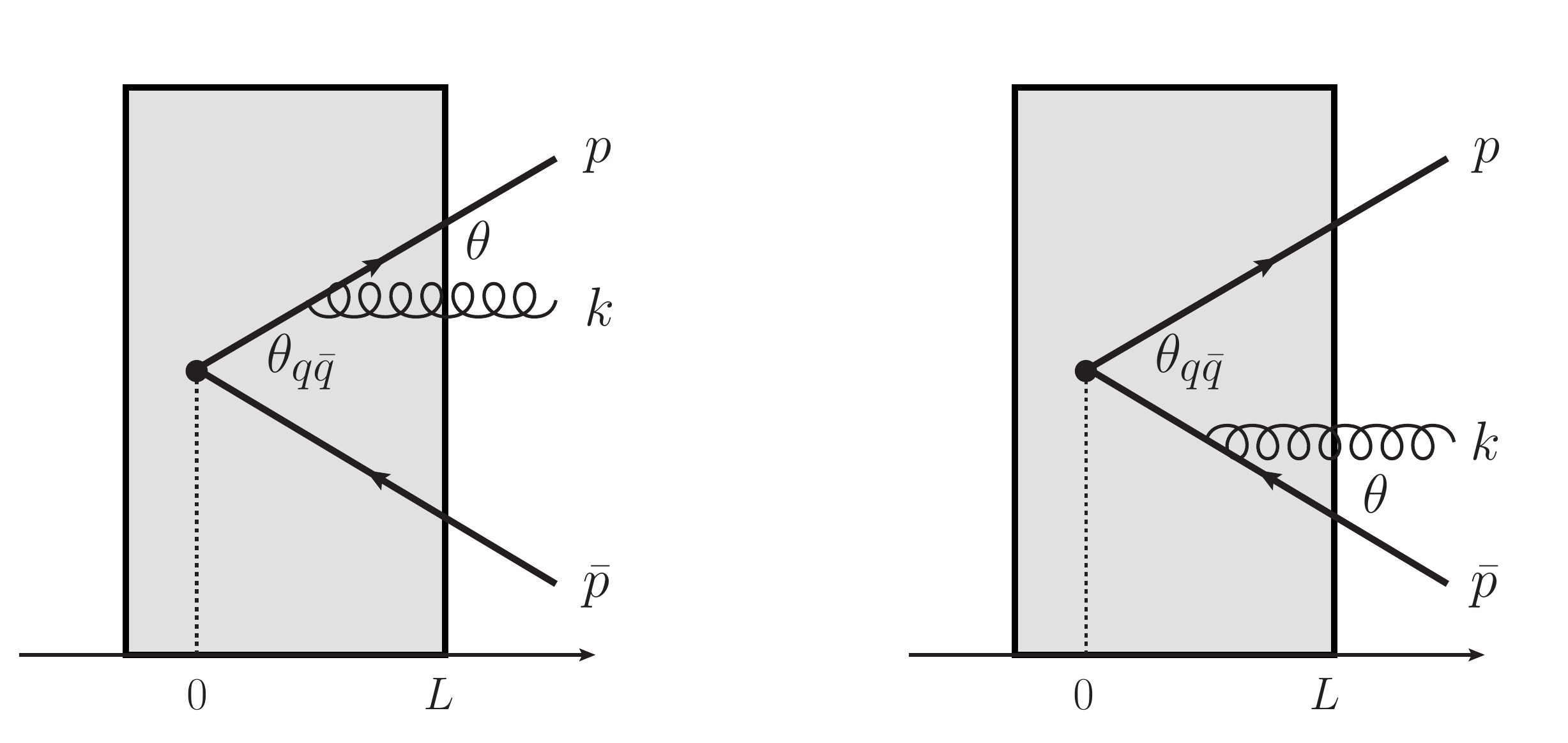}

\hspace{1.5cm}

\includegraphics[width=0.45\textwidth]{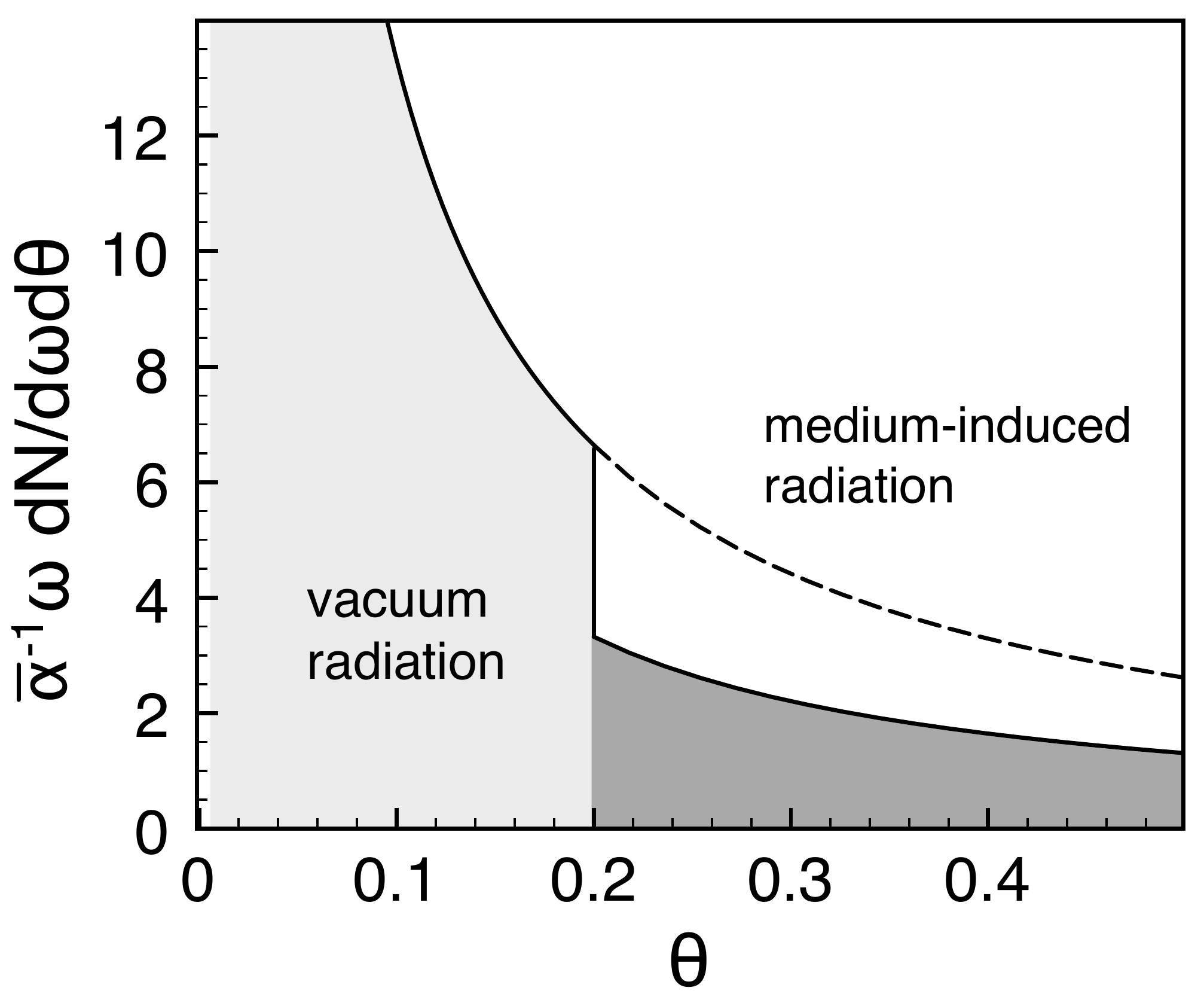}
\caption{(top) in-medium $q\bar{q}$ antenna (figure from \cite{MehtarTani:2011vh}) ; (bottom) angular distribution of the soft gluon radiation from a quark in an in-medium $q\bar{q}$ antenna with opening angle $\theta_{q\bar{q}}=0.2$ (figure from \cite{MehtarTani:2011tz}).}
\label{fig:ant}       
\end{figure}

Rigorous attempts to address this issue have been carried out recently \cite{MehtarTani:2010ma,MehtarTani:2011tz,MehtarTani:2011jw,MehtarTani:2011gf,CasalderreySolana:2011rz}. In a nutshell, momentum transfers and colour exchanges with the medium lead to  the decoherence of emitters and consequently to the breakdown of the familiar angular ordered branching pattern. In particular,  it was argued in \cite{MehtarTani:2010ma} that,  for a $q\bar{q}$ antenna, such decoherence effects result in medium induced radiation being anti-angular ordered, see fig.~\ref{fig:ant}. 

It should be noted, however, that these studies have  addressed so far only the radiation pattern of singlet and octet $q\bar{q}$ antennas, with the 
phenomenologically relevant case of emission from a $qg$ antenna yet to be considered. A further limitation of these studies is the assumption that the radiated gluon is much softer than the emitters.

Ongoing efforts \cite{aacm} to address the emission of two gluons, without the requirement of a relative softness hierarchy,  from a hard parton (quark or gluon) will yield a full set of jet calculus rules to be eventually implemented at the Monte Carlo level.

\section{Mass effects: heavy quarks}
\label{sec:heavy}

\begin{figure}
\includegraphics[width=0.45\textwidth]{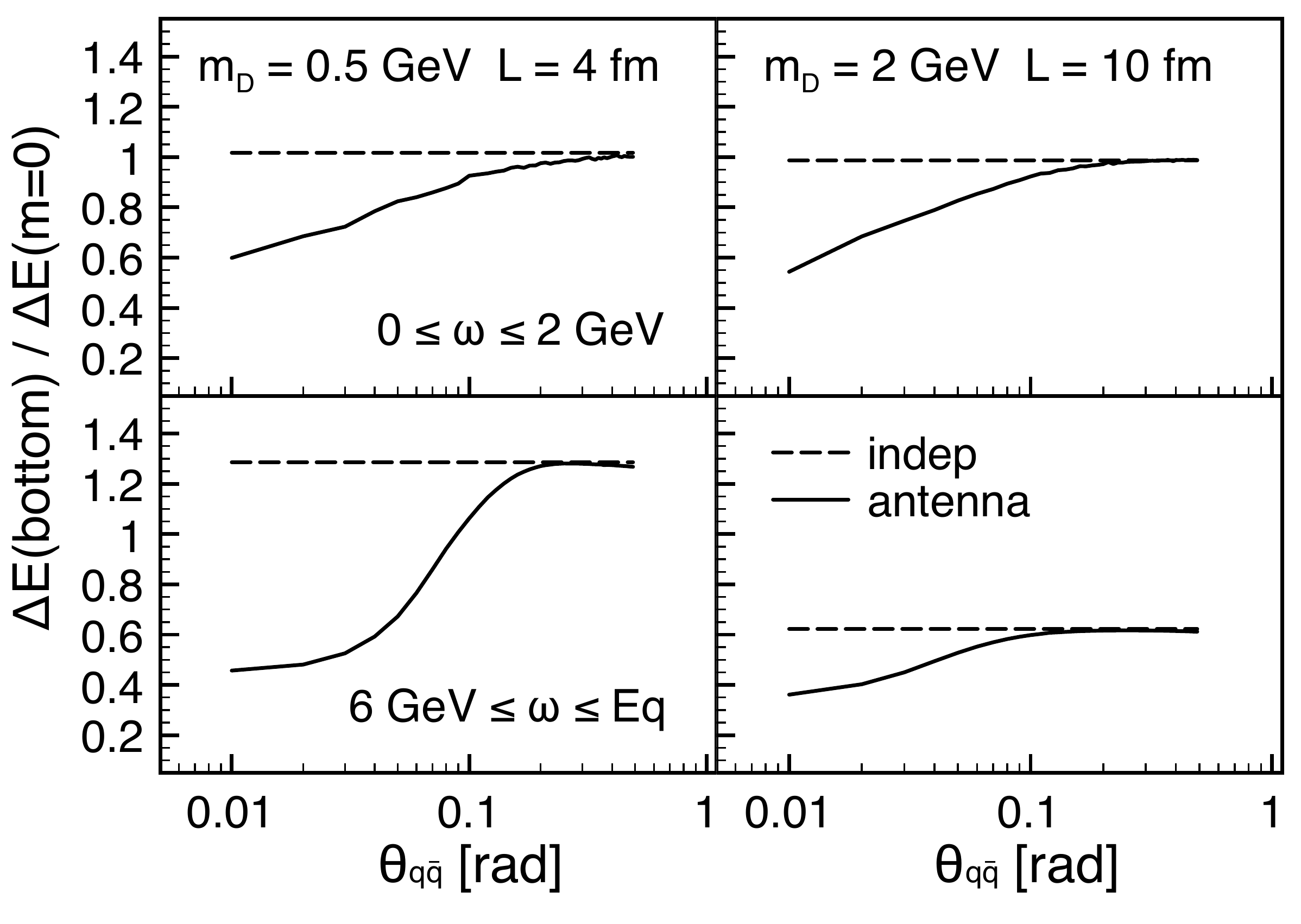}
\caption{energy loss of a b-quark relative to the massless case for both independent emission and antenna interference scenarios (figure from \cite{Armesto:2011ir}).}
\label{fig:heavy}       
\end{figure}

The energy loss of massive partons (heavy quarks) is expected to be smaller than in the massless case (light quarks and gluons). This is so due to a veto of radiation at small angle, the so-called dead cone effect. Given that medium induced decoherence effects lead to a similar suppression, the assessment of their combined net effect is of great interest. The antenna calculation (see sec.~\ref{sec:mult}) has been generalized to the massive case in \cite{Armesto:2011ir}. The result, in fig.~\ref{fig:heavy}, shows a clear difference between the case where emissions are independent and when the in-medium antenna interference pattern is accounted for.

\section{Jet observables}
\label{sec:jets}

The measurement of the dijet asymmetry $A_J= (E_{T_1}-E_{T_2})/(E_{T_1}+E_{T_2})$ in Pb-Pb collisions by both ATLAS \cite{Aad:2010bu,Cole:2011zz} and CMS \cite{Chatrchyan:2011sx,Roland:2011cs} provides a clear example of the extent to which such full jet measurements can contribute to furthering  the understanding of the dynamical processes responsible for jet quenching and highlights the insufficiencies of the existing theory to address such observables.

The measured event asymmetry distribution shows qualitative features consistent with a substantial medium induced energy loss of the recoiling jet. In a nutshell, the fraction of energy lost from the recoiling jet cone is larger in heavy ion collisions than in the proton-proton case, and grows with increasing centrality of those collisions. Crucially, this effect is accompanied by a distortion of the dijet azimuthal distribution which is, at most, very mild and centrality independent. 

\begin{figure}
\includegraphics[width=0.45\textwidth]{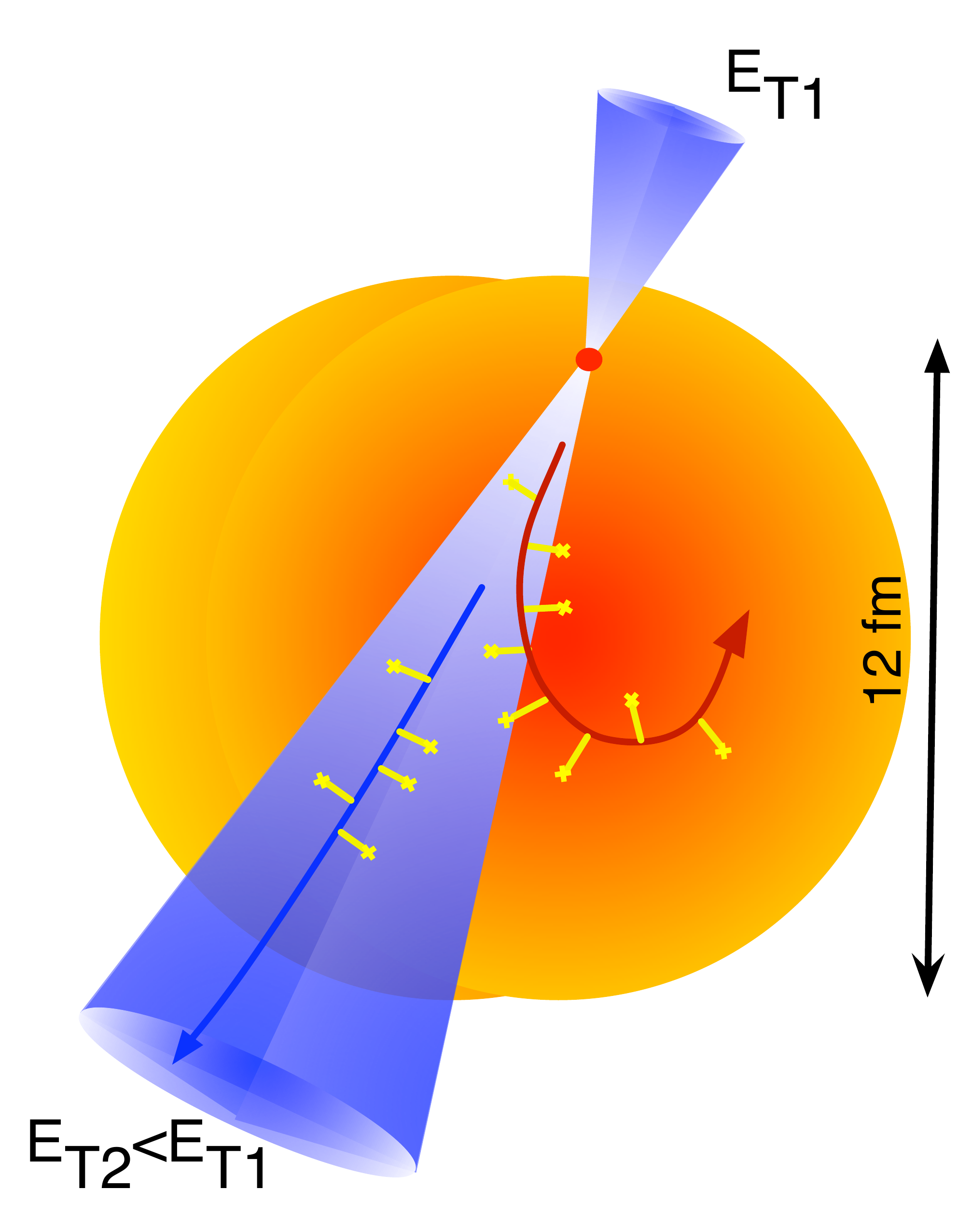}
\hspace{1.5cm}
\includegraphics[width=0.48\textwidth]{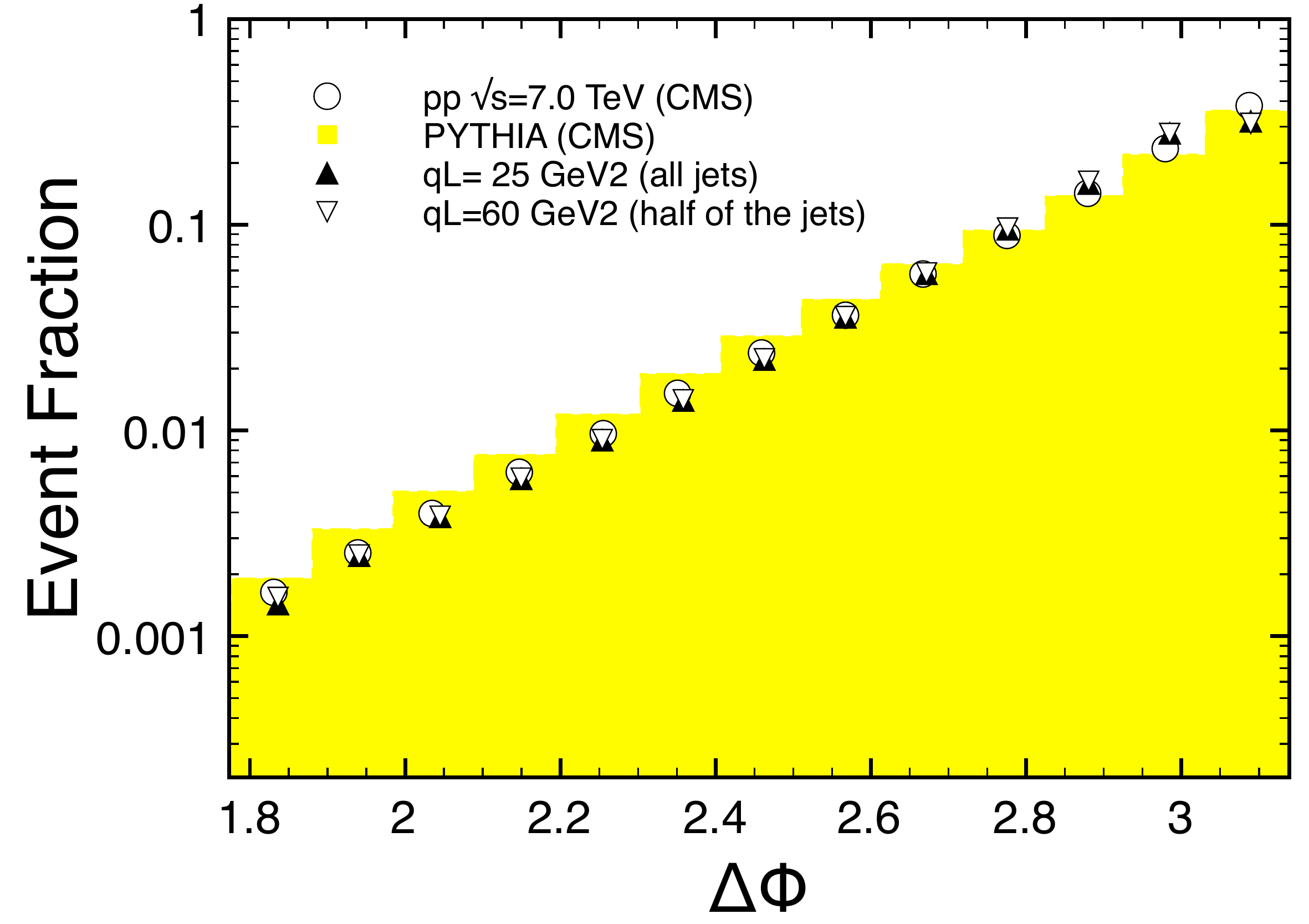}
\caption{(top) jet collimation: soft gluons emitted at small angles are transported outside the jet cone by their multiple random scatterings with medium components. (bottom) azimuthal angle distribution of dijet events: the considered values of $\hat q L$ correspond to the cases in which either all recoiling jets are taken as interacting with the medium or only a fraction ($\sim$ 0.5) do (figures from \cite{CasalderreySolana:2011rq}).}
\label{fig:jc}
\end{figure}

A novel mechanism --- jet collimation --- that leads to both large energy degradation and no sizeable azimuthal displacement, see fig.\,\ref{fig:jc} (bottom), was proposed in \cite{us}. Soft gluons radiated at small angles result in no azimuthal displacement of the jet. These gluons, and in fact all partons in the parton shower, undergo Brownian motion during their passage through the medium and thus accumulate an average squared transverse momentum $\hat{q}$ per unit path length, $\langle k_T^2 \rangle \sim \hat{q} L\sim \hat{q} \tau$ (see discussion in sec.~\ref{sec:jetcalc}). Thus, in the presence of a medium, the average formation time $\tau\sim \omega/k_T^2$ for partons of energy $\omega$ is 
\begin{equation}
	\langle \tau \rangle \sim \sqrt{\frac{\omega}{\hat{q}}} 
\end{equation}
As a result, soft jet fragments are formed early and those with energy $\omega\leq \sqrt{\hat{q} L}$ will be completely decorrelated from the initial jet direction, see fig.\,\ref{fig:jc} (top). In other words, the medium acts a frequency collimator, efficiently trimming away the soft components of the jet by transporting them to large angles. This is in agreement with the observation by CMS  \cite{Chatrchyan:2011sx,Roland:2011cs} that the energy lost from a jet cone is fully recovered at large angles and in the form of soft partons. It should be noted that this mechanism is effective for \textit{all} radiated soft quanta, in particular to vacuum-like emissions, and thus at play even in the absence of any medium induced radiation.

While the proposal in \cite{us} relies on simple heuristic arguments, a similar idea was developed in the context of a specific model in \cite{Qin:2010mn}. Importantly, Monte Carlo implementations \cite{Young:2011qx,Linders:2011qm} of jet quenching that account for collisional energy loss of the type advocated in \cite{us} have been showed to reproduce simultaneously the observed asymmetry and the non-modification of the azimuthal distribution.

The inference of microscopic dynamics from reconstructed jet data, tempting as it is, deserves a note of caution. Possible caveats on the reconstruction procedures employed to subtract the large and fluctuating background present in the heavy ion environment, and consequently on the robustness of the results, have been voiced in \cite{Cacciari:2011tm}, thus emphasizing the importance of detailed background studies as those conducted by the ALICE collaboration \cite{Abelev:2012ej}.

\begin{figure}
\begin{center}
\includegraphics[width=0.45\textwidth]{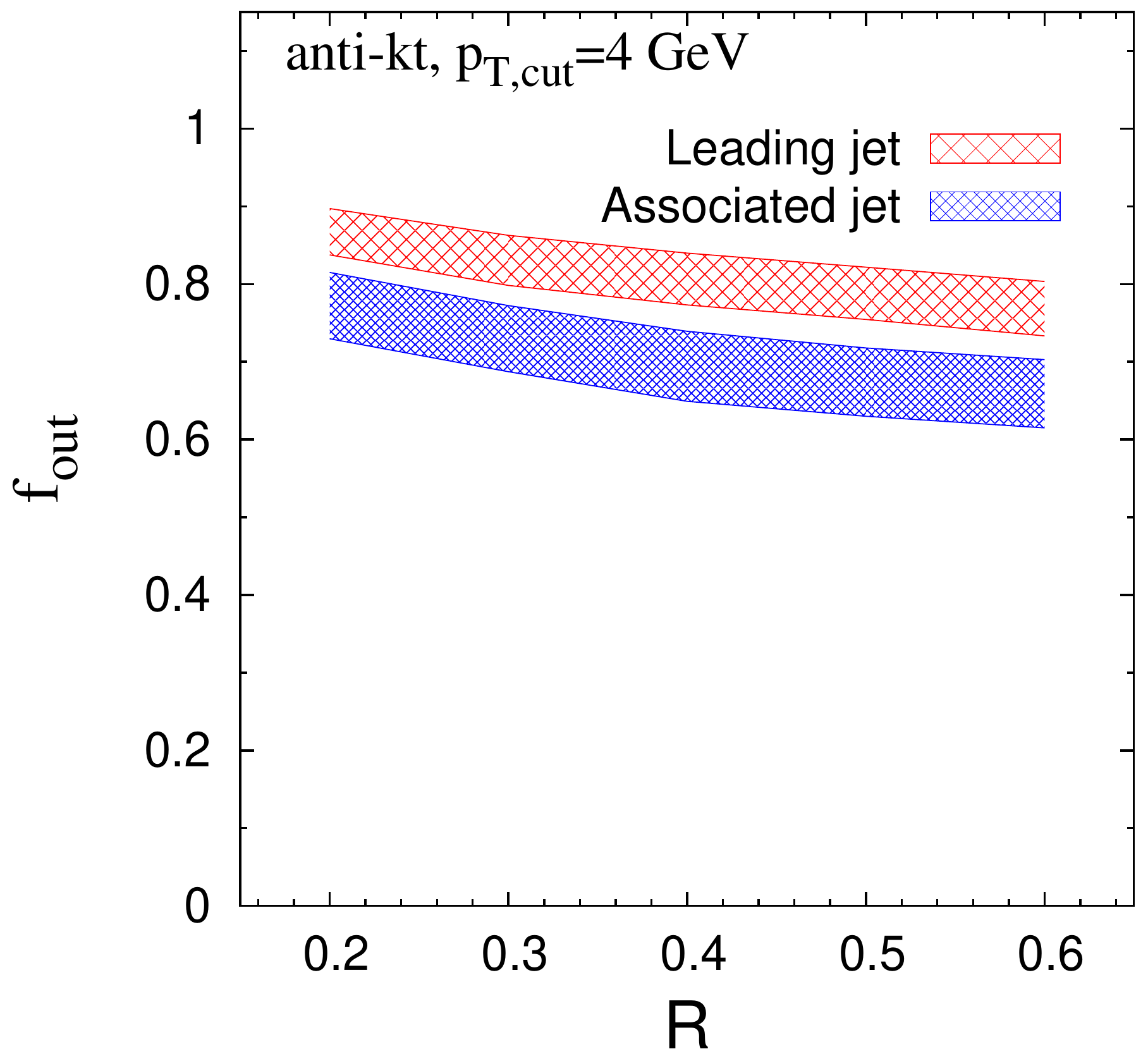}
\hspace{1.5cm}
\includegraphics[width=0.45\textwidth]{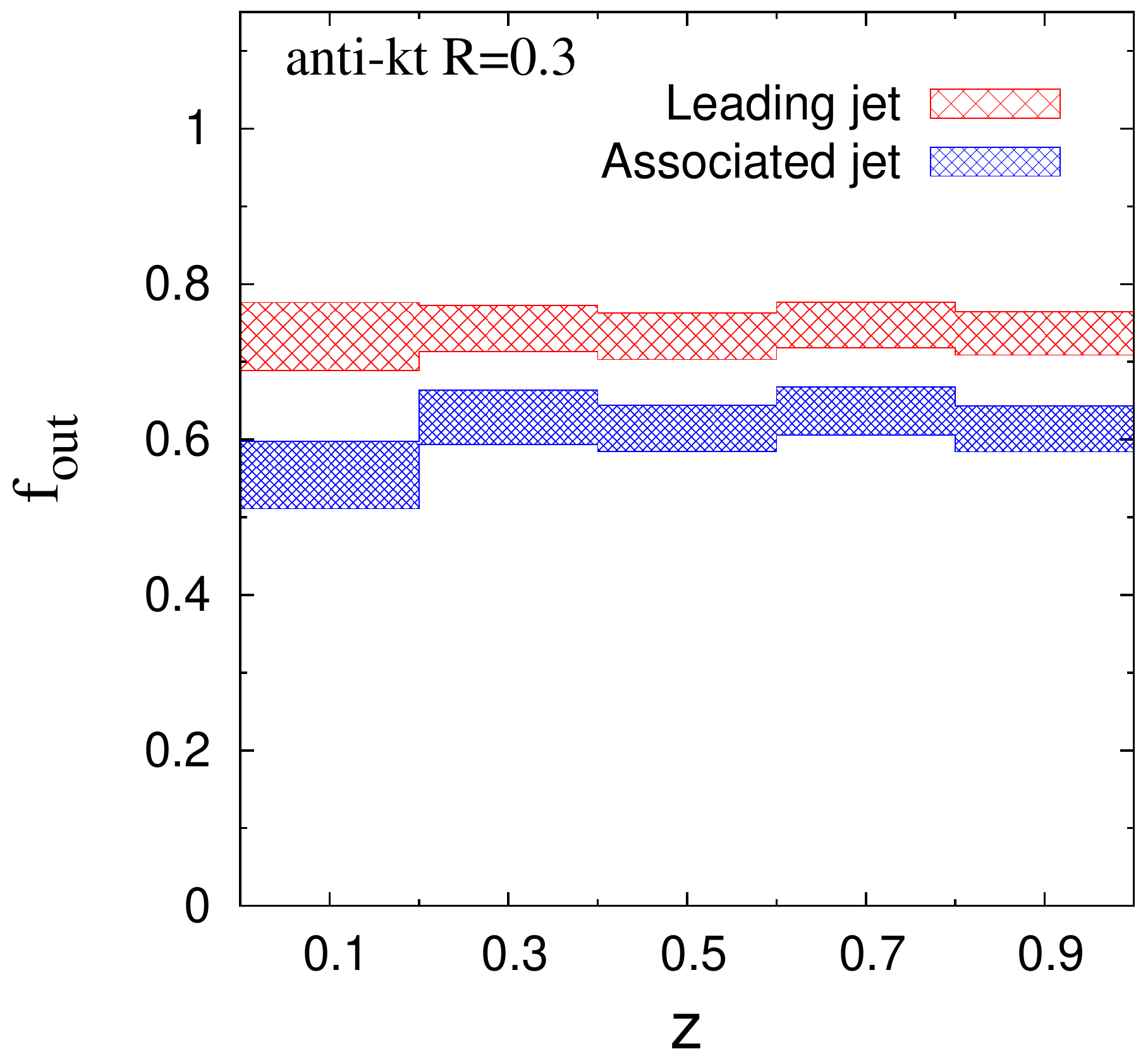}
\end{center}
\caption{Fraction of splittings of the parent parton that occur outside of the medium, assuming that in heavy ion collisions jet evolution proceeds as in vacuum. The analysis is performed for di-jet events of leading jet transverse momentum $p_{T,leading}>100$ GeV and associated jet momentum $p_{T,associated}> 40$ GeV and for a lead-lead collision with centrality class $0-30$ \%. A very mild dependence of this fraction 
is observed both in the reconstruction radius $R$ (top) and in the $z-$fraction of transverse momentum (bottom) (figure from \cite{CasalderreySolana:2011gx}).}
\label{fig:inout}
\end{figure}

Preliminary measurements of jet fragmentation functions, conducted by ATLAS \cite{Angerami:2011is}and CMS \cite{Collaboration:2011nsb} indicate that no deviation is observed, within errors, with respect to the pp case. These observations appear to be at odds with the overwhelming evidence for a sizable medium induced energy loss. A simple reconciliatory argument was presented in \cite{CasalderreySolana:2011gx}.There, a first attempt to estimate the time-scale for development of a jet was carried out. Relying on the estimate of the times at which branchings occur in \textit{vacuum} parton shower, it was shown that a significant fraction of splittings occur (see fig.~\ref{fig:inout}) outside a medium of realistic dimensions and geometry.

A contrasting reasoning, relying on the assumption that the entire jet develops and hadronizes within the medium, appeared in \cite{Loshaj:2011jx}. There, the authors argue within toy model discussion that non-perturbative interference effects lead to a scaling, and thus non-modification, of jet fragmentation functions.

Regardless of what the correct final answer will be, these studies clearly stress the need for a space-time formulation of a parton shower to be developed.

\section{Colour flow}
\label{sec:colour}

Standard energy loss calculations are performed colour-inclusively, that is to say the effects of colour flow between parton and medium are considered and the calculations are interfaced with hadronization procedures as if the final state partons had a vacuum colour distribution.
It was argued a few years ago that colour reconnections with the medium should result in modifications of the hydrochemistry of a jet  \cite{Sapeta:2007ad}.

Recently, a proof-of-principle study was performed \cite{Beraudo:2011bh} in which medium-induced colour flow was considered within a standard QCD calculation of parton energy loss. It results that colour exchanges with the medium can induce a characteristic softening of the ensuing hadronization regardless of where it happens.

\begin{figure}
\begin{center}
\includegraphics[width=0.45\textwidth]{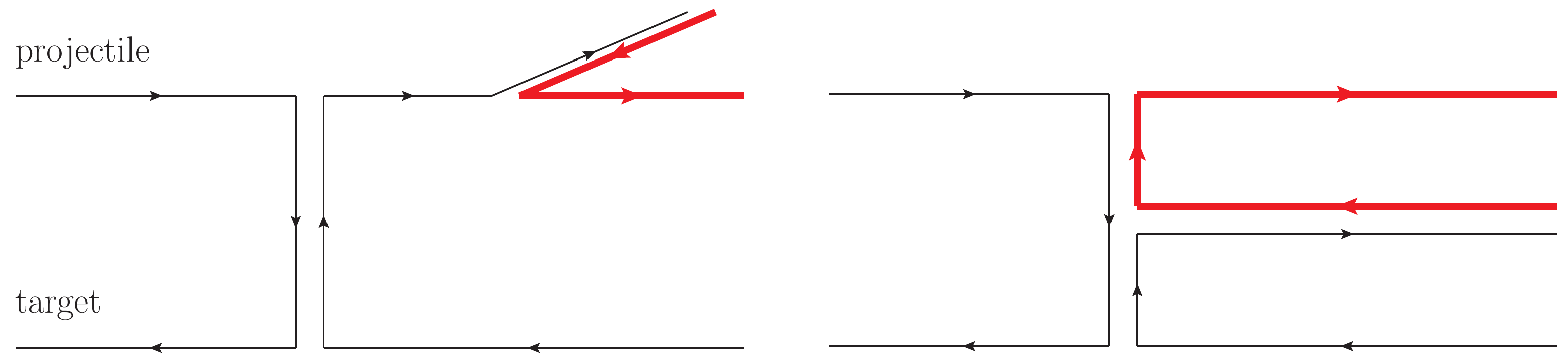}
\hspace{1.5cm}
\includegraphics[width=0.45\textwidth]{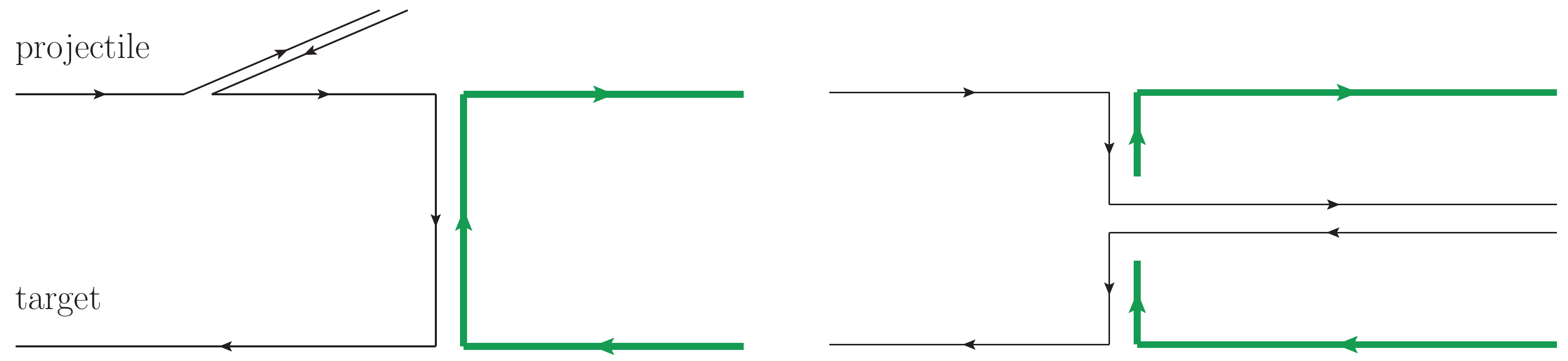}
\end{center}
\caption{$N=1$ opacity diagrams for gluon radiation from a projectile quark
in the large-$N_c$ limit. The most energetic color-singlet clusters are denoted
by thicker lines and correspond to color flow between projectile components
(top) contribution $dI^{\rm med}_{{\rm low}\, M}$) or between projectile 
and target  (bottom) contribution $dI^{\rm med}_{{\rm high}\, M}$). Diagrams on the right hand
side include a 3-gluon vertex (figure from  \cite{Beraudo:2011bh}).}
\label{fig:channels}
\end{figure}

This calculation, performed to leading order in opacity $\zeta\equiv L^+/\lambda^+_{\rm el}$ (in-medium 
path length $L^+$ and an elastic mean free path $\lambda^+_{\rm el}$), revisits the standard computation of medium-induced radiation spectrum 
by identifying contributions with distinct colour flow properties, see fig.~\ref{fig:channels}. In brief, contributions where the projectile parton remnant is colour connected to the target (fig.~\ref{fig:channels} (bottom)) will yield hadronizable configurations (colour singlet clusters or strings depending on the specific realistic implementation of the hadronization dynamics) of substantially larger invariant mass and consequently result in a softened hadronic final state.

The phenomenological consequences of the calculation were illustrated by assessing its effect on the simplest jet quenching observable, the nuclear modification factor $R_{AA}$.  The net result is a sizable (depending on the relative weight $f_t$ of the contributions with medium modified colour flow)  suppression of $R_{AA}$ persistent over a wide $p_t$ range, which can affect significantly the extraction of medium properties from the measured nuclear modification factor.

\begin{figure}
\begin{center}
\includegraphics[width=0.50\textwidth]{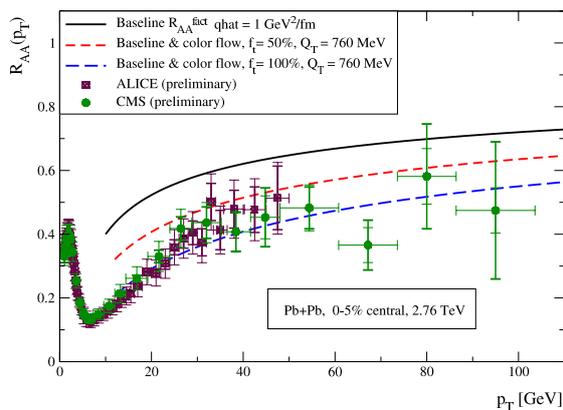}
\end{center}
\caption{The nuclear modification factor $R_{\rm AA}(p_T)$. The baseline calculation of kinematic
effects (solid black curve) is supplemented with the effect of color-flow 
modified hadronization (figure from  \cite{Beraudo:2011bh}). }
\label{fig:raa}
\end{figure}

Further, the colour decorrelation of final state partons due to medium induced colour exchanges has been found \cite{Aurenche:2011rd} to open additional colour channels, forbidden in the vacuum, and thus lead additional baryon production.

\section{Conclusions}

High-$p_t$ physics in heavy ion collisions is at present a very active field.  Several novel ideas have appeared recently and overall have reshaped the theory of jet quenching into a draft of a fully rigorous approach. More than ever it is clear that significant progress will rely on systematic and open-minded cooperation between the experimental and theoretical communities. 

As a final remark I would like to stress the importance of first principle Monte Carlo implementations of the microscopic dynamics, and their embedding in realistic simulations of the hot and dense QCD medium, as the single most important factor determining our ability to unravel comprehensively, for the first time, the  collective properties of a non-abelian gauge field theory that describes fundamental degrees of freedom of Nature.

\section*{Acknowledgments.}
I thank the conference organizers both for the invitation and for the exciting environment they created. I acknowledge the support of Funda\c c\~ao para a Ci\^encia e a Tecnologia (Portugal) under project CERN/FP/116379/2010

\end{document}